\documentclass[aps,preprint]{revtex4}%
\usepackage{amsfonts}
\usepackage{amsmath}
\usepackage{amssymb}
\usepackage{subfigure}
\usepackage{graphicx}%

\begin{document}
\title{Thermodynamics with pressure and volume of 4D Gauss-Bonnet AdS Black Holes under the scalar field}
\author{Benrong Mu$^{a,b}$}
\email{benrongmu@cdutcm.edu.cn}

\author{Jing Liang$^{a,b}$}
\email{jingliang@stu.scu.edu.cn}

\author{Xiaobo Guo$^{c}$}
\email{guoxiaobo@czu.edu.cn}

\affiliation{$^{a}$Physics Teaching and Research section, College of Medical Technology,
Chengdu University of Traditional Chinese Medicine, Chengdu, 611137,
PR China}
\affiliation{$^{b}$Center for Theoretical Physics, College of Physics, Sichuan University, Chengdu, 610064, PR China}
\affiliation{$^{c}$Mechanical and Electrical Engineering School, Chizhou University, Chizhou, 247000, PR China}

\begin{abstract}
By using the scattering of a scalar field, we discuss the thermodynamics and overcharging problem in a 4D Gauss-Bonnet AdS black hole in both the normal phase space and extended phase space. In the normal phase space, where the cosmological constant and Gauss-Bonnet parameter are fixed, the first law and the second law of thermodynamics are valid. In addiction, the black hole cannot be overcharged and the weak cosmic censorship conjecture is valid. In the extended phase space, where the cosmological constant and Gauss-Bonnet parameter are treated as the thermodynamic variables, the first law is still valid. However, the second law is indefinite. Moreover, after the scattering of the scalar field, the extremal black hole cannot be overcharged and the near-extremal black hole can be overcharged.

\end{abstract}

\maketitle
\tableofcontents{}

\bigskip{}



\section{Introduction}
\label{sec:intro}
It is well known that space-time singularities can be form at the end of gravitational collapse. Near the singularity, the laws of physics fail. In order to avoid physical damage caused by singularities, Penrose proposed the weak cosmic censorship conjecture (WCCC). It points out that under normal material properties and initial conditions, naked singularities cannot be formed in real physical processes, and the singularities are covered by the event horizon \cite{intro-Penrose:1969pc}. The WCCC plays an important part in black hole physics, and its validity is a necessary condition to ensure the predictability of the laws of physics. Although its correctness has been widely accepted, there is no complete evidence to prove its correctness. People can only test its correctness in different space-time. The Gedanken experiment designed by Wald is an effective method to test the validity of WCCC \cite{intro-Sorce:1974dst}. In this experiment, a particle with sufficient charge or angular momentum is thrown into a charged black hole to test the evolution of the black hole. If the final state of the black hole evolution is no longer a black hole, then the event horizon does not exist and the singularity is naked. According to Wald's work, the effectiveness of WCCC has been tested in various space-time \cite{intro-Ying:2020bch, intro-Yang:2020czk, intro-Mu:2019bim, intro-Hu:2019lcy, intro-He:2019fti, intro-Wang:2019dzl, intro-Liu:2020cji, intro-Hu:2020lkg, intro-Wang:2020osg, intro-Wang:2019jzz, intro-Chen:2020zps, intro-Hu:2020ccj, intro-Zeng:2019baw, intro-Zeng:2019hux, intro-Zeng:2019aao, intro-Han:2019lfs, intro-Han:2019kjr, intro-Zeng:2019jrh}. In addition, Semiz first used a complex scalar field to replace particles to test the effectiveness of WCCC, and found that the strong Kerr-Newman black hole cannot be overcharged \cite{intro-Semiz:2005gs}. There are also many papers that have verified the effectiveness of WCCC in different black holes through this method \cite{intro-Hong:2020zcf,intro-Yang:2020iat,intro-Bai:2020ieh,intro-Chen:2019nsr,intro-Chen:2018yah,intro-Shaymatov:2019del,intro-Gwak:2018akg,intro-Toth:2011ab,intro-Goncalves:2020ccm,intro-Jiang:2020btc,intro-Duztas:2018adf,intro-Duztas:2019mxr}. In particular, RN-AdS black hole is studied in both normal and extended phase space \cite{intro-Gwak:2017kkt, intro-Gwak:2019asi}. For RN-AdS black holes, the first law of thermodynamics and WCCC are always satisfied, while the second law of thermodynamics is valid in normal phase space and violated in the extended phase space. Moreover, the thermodynamics and WCCC of the high-dimensional Gauss-Bonnet AdS black hole $(d\geq5)$ are studied in \cite{intro-Zeng:2019hux}.

Generally speaking, no matter is able to escape through the black hole's event horizon. So that external observers cannot detect any radiation from inside the black hole. From the perspective of quantum physics, there is quantum tunnelling in black holes. From this, the temperature of the black hole can be defined, and the black hole can be regarded as a thermal system with Hawking temperature \cite{intro-Hawking:1974sw, intro-Hawking:1976de}. In addition, black holes have irreducible mass, which is a property that increases with irreversible processes \cite{intro-Christodoulou:1970wf, intro-Christodoulou:1972kt, intro-Smarr:1972kt, intro-Bardeen:1970zz}. The irreducible mass is similar to the entropy in the thermal system and based on this similarity, the entropy of the black hole can be obtained. This entropy is the Bekenstein-Hawking entropy of the black hole \cite{intro-Bekenstein:1973ur, intro-Bekenstein:1974ax}, which is proportional to the area of the horizon. Using these two thermodynamic properties, temperature and entropy, the laws of thermodynamics for black holes as a thermal system is established. The first law of black hole thermodynamics is usually written as
\begin{equation}
dM=TdS+\varPhi dQ.
\label{eqn:dM1}
\end{equation}
Where $M$ is the mass, $T$ is the Hawking temperature, $S$ is the entropy, $\varPhi$ is the electric potential and $Q$ is the electric charge. The mass is usually interpreted as the enthalpy \cite{intro-Kastor:2009wy}. It is worth noting that there is no $PdV$ term in Eq. $\left(\ref{eqn:dM1}\right)$. Recently, an interesting idea has been proposed. The cosmological constant can be explained by thermodynamic pressure. When the cosmological constant, $\varLambda$, is treated as the pressure of the black hole \cite{intro-Dolan:2011xt,intro-Kubiznak:2012wp,intro-Cvetic:2010jb,intro-Caceres:2015vsa,intro-Hendi:2012um,intro-Pedraza:2018eey} and the volume of the black hole is defined as the thermodynamic variable conjugate to the pressure \cite{intro-Dolan:2010ha}, Eq. $\left(\ref{eqn:dM1}\right)$ is modified as
\begin{equation}
dM=TdS+VdP+\varPhi dQ.
\label{eqn:dM2}
\end{equation}
where $P$, $\varLambda$ and $V$ satisfy $P=-\frac{\varLambda}{8\pi}$, $V=\left(\frac{\partial M}{\partial P}\right)_{S,Q}$.

As we all know, there are static spherically symmetric black hole solutions with $d\geq5$ in Gauss-Bonnet gravity \cite{intro-Boulware:1985wk,intro-Wiltshire:1985us,intro-Cai:2001dz,intro-Nojiri:2001aj,intro-Cvetic:2001bk}. When $d=4$, the Gauss-Bonnet term becomes a topological invariance and does not contribute to the field equation in the four-dimensional space-time, so there is no 4D Gauss-Bonnet black hole. However, in the recent work of Glavan and Lin \cite{intro-Glavan:2019inb}, it is proposed that the non-trivial solution of the four-dimensional black hole can be obtained by re-adjusting the Gauss-Bonnet coupling parameter $\alpha$ to $\alpha\rightarrow\alpha/\left(d-4\right)$, and then taking the limit of $d\rightarrow4$. Subsequently, the black hole solution is extended to charged case in an anti-de Sitter (AdS) space \cite{intro-Fernandes:2020rpa}. The first law of black hole thermodynamics of 4D Gauss-Bonnet AdS black hole is
\begin{equation}
dM=TdS+\varphi dQ+VdP+\mathcal{A}d\alpha.
\end{equation}
where $\mathcal{A}$ is the conjugate quantity of Gauss-Bonnet parameter and is defined as
\begin{equation}
\mathcal{A}=\left(\frac{\partial M}{\partial\alpha}\right)_{S,Q,P}=\frac{1}{2r_{+}}+2\pi T\left(1-2ln\frac{r_{+}}{\sqrt{\alpha}}\right).
\end{equation}
When taking the limit $\alpha\rightarrow0$, the RN-AdS will be recovered. There are many studies on the thermodynamic properties of 4D Gauss-Bonnet black holes \cite{intro-EslamPanah:2020hoj, intro-Hegde:2020xlv, intro-Wei:2020poh, intro-HosseiniMansoori:2020yfj, intro-Singh:2020xju, intro-Konoplya:2020bxa, intro-Konoplya:2020juj, intro-Li:2020tlo, intro-Zhang:2020sjh, intro-Mishra:2020gce}.

The present paper is organized as follows. In Sec. \ref{sec:M}, we briefly review 4D Gauss-Bonnet AdS black hole solution. In Sec. \ref{sec:E and Q}, we investigate the complex scalar field in the black hole background. Next, we briefly discuss the first law and the second law of thermodynamics of the black hole in the normal and extended phase space in Sec. \ref{sec:Thermodynamics}. Furthermore, the overcharging problem in the normal and extended phase space is discussed in Sec. \ref{sec:wccc}. Sec. \ref{sec:con} is devoted to our discussion and conclusion..

\section{4D Gauss-Bonnet AdS black holes}
\label{sec:M}
\begin{figure}[htb]
\centering
\includegraphics[scale=0.8]{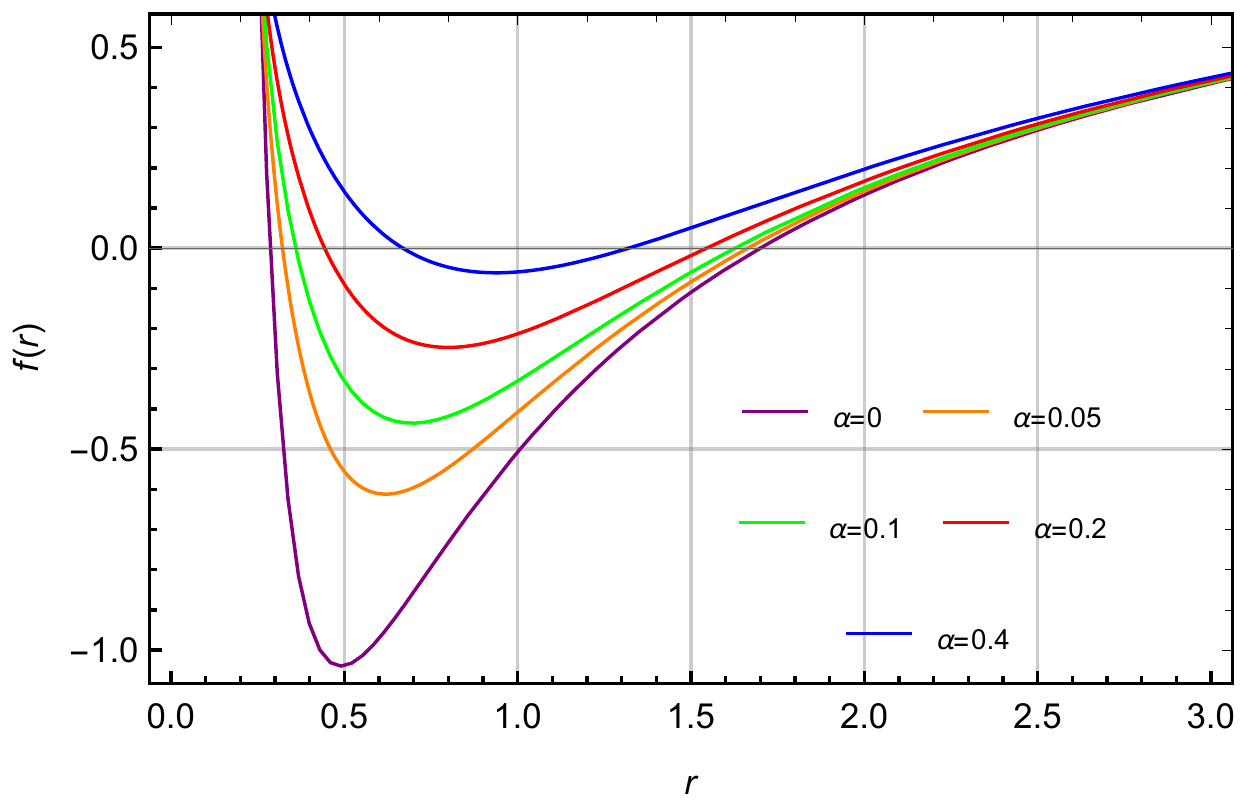}
\caption{The function $f(r)$ for different values of $\alpha$. We choose $M=1$ and $Q=0.7$.}
\label{fig:f1}
\end{figure}

The action of the Gauss-Bonnet-Maxwell theory in a d-dimensional space-time is \cite{intro-Glavan:2019inb}
\begin{equation}
S=\frac{1}{16\pi}\int d^{D}x\sqrt{-g}\left[R-2\varLambda+\frac{\alpha}{d-4}\left(r^{2}-4R_{\mu\nu}R^{\mu\nu}+R_{\mu\nu\rho\sigma}R^{\mu\nu\rho\sigma}\right)-F^{\mu\nu}F_{\mu\nu}\right],
\end{equation}
where $\Lambda$ is the cosmological constant that relates to the
AdS radius $l$ with the relation $\Lambda=-\frac{\left(D-1\right)\left(D-2\right)}{2l^{2}}$, $g$ is determinant of the metric tensor and $F_{\mu\nu}$ is the Maxwell field strength.
Adopting the limit $d\rightarrow4$ and solving the field equations, we obtain
\begin{equation}
ds^{2}=f(r)dt^{2}-\frac{1}{f(r)}dr^{2}-r^{2}\left(d\theta^{2}+sin^{2}\theta d\phi^{2}\right),
\end{equation}
\begin{equation}
f\left(r\right)=1+\frac{r^{2}}{2\alpha}\left[1-\sqrt{1+4\alpha\left(-\frac{1}{l^{2}}+\frac{2M}{r^{3}}-\frac{Q^{2}}{r^{4}}\right)}\right],
\end{equation}
where $M$ and $Q$ are the mass and electric charge of the black hole, respectively. $\alpha$ is the Gauss-Bonnet coupling constant. The metric function is written as
\begin{equation}
f\left(r\right)=\frac{2\alpha+2r^{2}-2\left(-\frac{r^{4}}{l^{2}}+2Mr-Q^{2}\right)}{2\alpha+r^{2}+\sqrt{r^{4}+4\alpha\left(-\frac{r^{4}}{l^{2}}+2Mr-Q^{2}\right)}},
\end{equation}
In Fig. \ref{fig:f1}, we plot the metric function $f(r)$ for different
situations. It can be observed from the figure that as the value of $\alpha$ increases, the value of $f(r)$ also increases. In the limit $\alpha\rightarrow0$, the RN-AdS black hole solution will be recovered.
Solving the equation $f(r) = 0$, we can obtain two solutions $r_-$ and $r_+$. The two solutions correspond to the radius of the Cauchy horizon and the event horizon. At the event horizon, the black hole mass is expressed as
\begin{equation}
M=\frac{\text{\ensuremath{\alpha}l}^{2}+l^{2}Q^{2}+l^{2}r_{+}^{2}+r_{+}^{4}}{2l^{2}r_{+}}.
\end{equation}
The Hawking temperature of the black hole is \cite{intro-Fernandes:2020rpa}
\begin{equation}
T=\frac{f^{\prime}(r_{+})}{4\pi}=\frac{-\alpha+r_{+}^{2}+3\frac{r_{+}^{4}}{l^{2}}-Q^{2}}{4\pi\left(r_{+}^{3}+2\alpha r_{+}\right)}.
\label{eqn:T}
\end{equation}
The electric potential and the entropy take on the forms as
\begin{equation}
\varphi=\frac{Q}{r_{+}},
\label{eqn:potential}
\end{equation}
\begin{equation}
S=\int\frac{dM}{T}=\pi r_{+}^{2}+4\pi\alpha ln\frac{r_{+}}{\sqrt{\alpha}}.
\label{eqn:entropy}
\end{equation}
In the normal phase space, where the cosmological constant is fixed, the first law of thermodynamics takes on the form as
\begin{equation}
dM=TdS+\varphi dQ.
\end{equation}
In the extended phase space, we construct the first law of thermodynamics by considering the cosmological constant as dynamic pressure, and its conjugate as thermodynamic volume. In this case, the black hole mass $M$ is interpreted as the enthalpy $H$ rather than the internal energy $U$ of the system. In addition, since the cosmological constant is regarded as a thermodynamic pressure, the Smarr relationship of black hole thermodynamics can be obtained through the scaling argument. In the Gauss-Bonnet gravity, in order to satisfy the Smarr relationship, the Gauss-Bonnet coefficient should also be regarded as a dynamic quantity. Therefore, the form of the first law in extended phase space is \cite{M-Cai:2013qga,M-Mahish:2019tgv}
\begin{equation}
dM=TdS+\varphi dQ+VdP+\mathcal{A}d\alpha,
\end{equation}
where $P$ is the pressure, $V$ is the conjugate of $P$, which is interpreted as volume, and $\mathcal{A}$ is the
conjugate quantity of Gauss-Bonnet coefficient $\alpha$, which are defined as \cite{intro-Wei:2020poh}
\begin{equation}
P=-\frac{\varLambda}{8\pi}=\frac{3}{8\pi l^{2}},
\end{equation}
\begin{equation}
V=\left(\frac{\partial M}{\partial P}\right)_{S,Q,\alpha}=\frac{4\pi r_{+}^{3}}{3},
\label{eqn:V}
\end{equation}
\begin{equation}
\mathcal{A}=\left(\frac{\partial M}{\partial\alpha}\right)_{S,Q,P}=\frac{1}{2r_{+}}+2\pi T\left(1-2ln\frac{r_{+}}{\sqrt{\alpha}}\right).
\label{eqn:A}
\end{equation}
The following Smarr relation is also satisfied
\begin{equation}
H=2TS+\varphi Q-2PV+2\mathcal{A}\alpha.
\label{eqn:M}
\end{equation}
\section{Energy and charge's variation of the 4D Gauss-Bonnet AdS black hole}
\label{sec:E and Q}
The charged complex scalar field $\psi$ with mass $m$ and charge $q$ satisfies
\begin{equation}
\left(\nabla^{\mu}-iqA^{\mu}\right)(\nabla_{\mu}-iqA_{\mu})\psi-m^{2}\psi=0,
\label{eqn:1}
\end{equation}
which can be written as
\begin{equation}
\frac{1}{\sqrt{-g}}\left(\partial_{\mu}-iqA_{\mu}\right)\left[\sqrt{-g}g^{\mu\nu}\left(\partial_{\nu}-iqA_{\nu}\right)\psi\right]-m^{2}\psi=0.
\end{equation}
Since space-time is static and spherically symmetric, the complex scalar field can be decomposed into
\begin{equation}
\psi=e^{-i\omega t}R(r)\varPhi(\theta,\phi),
\label{eqn:psi}
\end{equation}
where $\omega$ is the energy of the particle, $R(r)$ is the radial function and $\varPhi(\theta,\phi)$ is spherical harmonic function.
As usual, we introduce the tortoise coordinate to solve the radial equation
\begin{equation}
\frac{dr}{dr_{\ast}}=f.
\end{equation}
Then, the radial equation in Eq. $\left(\ref{eqn:psi}\right)$ is rewritten as
\begin{equation}
R(r)=e^{\pm i\left(\omega-\frac{qQ}{r}\right)r_{\ast}},
\end{equation}
where $+/-$ corresponds to the solution of the outgoing/ingoing radial wave. Since we discuss the thermodynamics and validity of the WCCC by scattering of the ingoing waves at the event horizon, we focus on the ingoing wave function.
The energy-momentum tensor is given by
\begin{equation}
T_{\nu}^{\mu}=\frac{1}{2}[\left(\partial^{\mu}-iqA^{\mu}\right)\psi^{\ast}\partial_{\nu}\psi+\left(\partial^{\mu}+iqA^{\mu}\right)\psi\partial_{\nu}\psi^{\ast}]+\delta_{\nu}^{\mu}\mathcal{L}.
\label{eqn:T nu mu}
\end{equation}
From Eq. $\left(\ref{eqn:T nu mu}\right)$, we can get the energy flux through the event horizon is
\begin{equation}
\frac{dE}{dt}=\int T_{t}^{r}\sqrt{-g}d\theta d\phi=\omega(\omega-q\varphi)r_{+}^{2}.
\end{equation}
In addition, the charge flux can get from the energy flux \cite{E-Bekenstein:1973mi}. Therefore, the charge flux is written as
\begin{equation}
\frac{dQ}{dt}=-\int j^{r}\sqrt{-g}d\theta d\varphi=q\left(\omega-q\varphi\right)r_{+}^{2}.
\end{equation}
According to the conservation of energy and charge, when the complex scalar field is scattered by the black hole, the decrease in the energy and charge of the scalar field is equal to the increase in the energy and charge of the black hole. In an infinitely time interval $dt$, the changes in the mass and charge of the black hole are
\begin{equation}
dU=dE=\omega(\omega-q\varphi)r_{+}^{2}dt,dQ=q(\omega-q\text{\ensuremath{\varphi}})r_{+}^{2}dt.
\label{eqn:dUdQ}
\end{equation}

\section{Thermodynamics under the scalar field}
\label{sec:Thermodynamics}
In this section, we investigate the black hole thermodynamics of the Einstein-Gauss-Bonnet gravity coupled to the
Maxwell theory in the normal and extended phase spaces by the scattering of a scalar field. For the convenience of the later discussion, before the subsequent discussions on the thermodynamic properties, we first give the following formulas
\begin{equation}
\begin{aligned}
& \frac{\partial f}{\partial M}|_{r=r_{+}}=-\frac{2}{r_{+}+\frac{2\alpha}{r_{+}}},\\
& \frac{\partial f}{\partial Q}|_{r=r_{+}}=\frac{2\varphi}{r_{+}+\frac{2\alpha}{r_{+}}},\\
& \frac{\partial f}{\partial l}|_{r=r_{+}}=-\frac{2r_{+}^{2}}{l^{3}}\frac{1}{1+\frac{2\alpha}{r_{+}^{2}}},\\
& \frac{\partial f}{\partial\alpha}|_{r=r_{+}}=\frac{1}{r_{+}^{2}+2\alpha},\\
& \frac{\partial f}{\partial r}|_{r=r_{+}}=4\text{\ensuremath{\pi}}T.
\end{aligned}
\label{eqn:T Df}
\end{equation}

\subsection{Thermodynamics in the normal phase space}
In normal phase space, cosmological constant is fixed, and black holes are characterized by mass $M$ and charge $Q$. During the scattering of the scalar field, the mass $M$, the charge $Q$ and other thermodynamic variables of the black bole change due to the conservation law. Assuming that the black hole's initial state is expressed by $(M,Q,r_{+})$
and final state is expressed by $(M+dM,Q+dQ,r_{+}+dr_{+})$. The initial state $(M,Q,r_{+})$ and the final state $(M+dM,Q+dQ,r_{+}+dr_{+})$ satisfy
\begin{equation}
f\left(M,Q,r_{+}\right)=f\left(M+dM,Q+dQ,r_{+}+dr_{+}\right)=0.
\label{eqn:nor T f}
\end{equation}
The functions $f\left(M,Q,r_{+}\right)$ and $f\left(M+dM,Q+dQ,r_{+}+dr_{+}\right)$ satisfy
the following relation
\begin{equation}
\begin{aligned}
&f\left(M+dM,Q+dQ,r_{+}+dr_{+}\right)=f\left(M,Q,r_{+}\right)\\
&+\frac{\partial f}{\partial M}|_{r=r_{+}}dM+\frac{\partial f}{\partial Q}|_{r=r_{+}}dQ+\frac{\partial f}{\partial r}|_{r=r_{+}}dr_{+},\\
\end{aligned}
\label{eqn:nor T f1 eqn}
\end{equation}
Inserting Eqs. $\left(\ref{eqn:T Df}\right)$ and $\left(\ref{eqn:nor T f}\right)$ to Eq. $\left(\ref{eqn:nor T f1 eqn}\right)$, we have
\begin{equation}
dM=2\pi T\left(r_{+}+\frac{2\alpha}{r_{+}}\right)dr_{+}+\varphi dQ.
\end{equation}
Considering Eqs. $\left(\ref{eqn:entropy}\right)$, the above equation is modified as
\begin{equation}
dM=TdS+\varphi dQ,
\label{eqn:nor 1st}
\end{equation}
which is the first law of thermodynamics.

In the normal phase space, the mass $M$ is interpreted as the internal energy of the thermodynamic system. Then, the changes in the internal energy and charge of the black hole within an infinitesimal time interval $dt$ is
\begin{equation}
dM=dU=\omega(\omega-q\varphi)r_{+}^{2}dt,dQ=q(\omega-q\text{\ensuremath{\varphi}})r_{+}^{2}dt.
\label{eqn:dMdQ}
\end{equation}
Using Eqs. $\left(\ref{eqn:nor 1st}\right)$ and $\left(\ref{eqn:dMdQ}\right)$, we have
\begin{equation}
dS=\frac{r_{+}^{2}(\omega-q\varphi)^{2}}{T}dt\geq0,
\label{eqn:nor T dS}
\end{equation}
which shows that the entropy of the black hole increases. Therefore, the second law of black hole thermodynamics is satisfied for the black hole in the normal phase space.

\subsection{Thermodynamics in the extended phase space}
In the extended phase space, the cosmological constant is treated as the function of the pressure
of the black hole. Moreover, the Gauss-Bonnet coupling constant $\alpha$ is also treated as a new thermodynamic variable \cite{T-Hu:2019lcy, T-Wei:2014hba}. The mass $M$ is regarded as the enthalpy $H$ of the thermodynamic system, not the thermodynamic energy $U$. The relationship between thermodynamic energy and enthalpy is \cite{T-Haldar:2019rxl, T-Lan:2018nnp}
\begin{equation}
M=U+PV.
\end{equation}
In this case, the variations of the energy and charge are
\begin{equation}
dU=d(M-PV)=\omega\left(\omega-q\varphi\right)r_{+}^{2}dt,dQ=q\left(\omega-q\varphi\right)r_{+}^{2}dt.
\end{equation}

\begin{figure}[htb]
\centering
\includegraphics[scale=0.8]{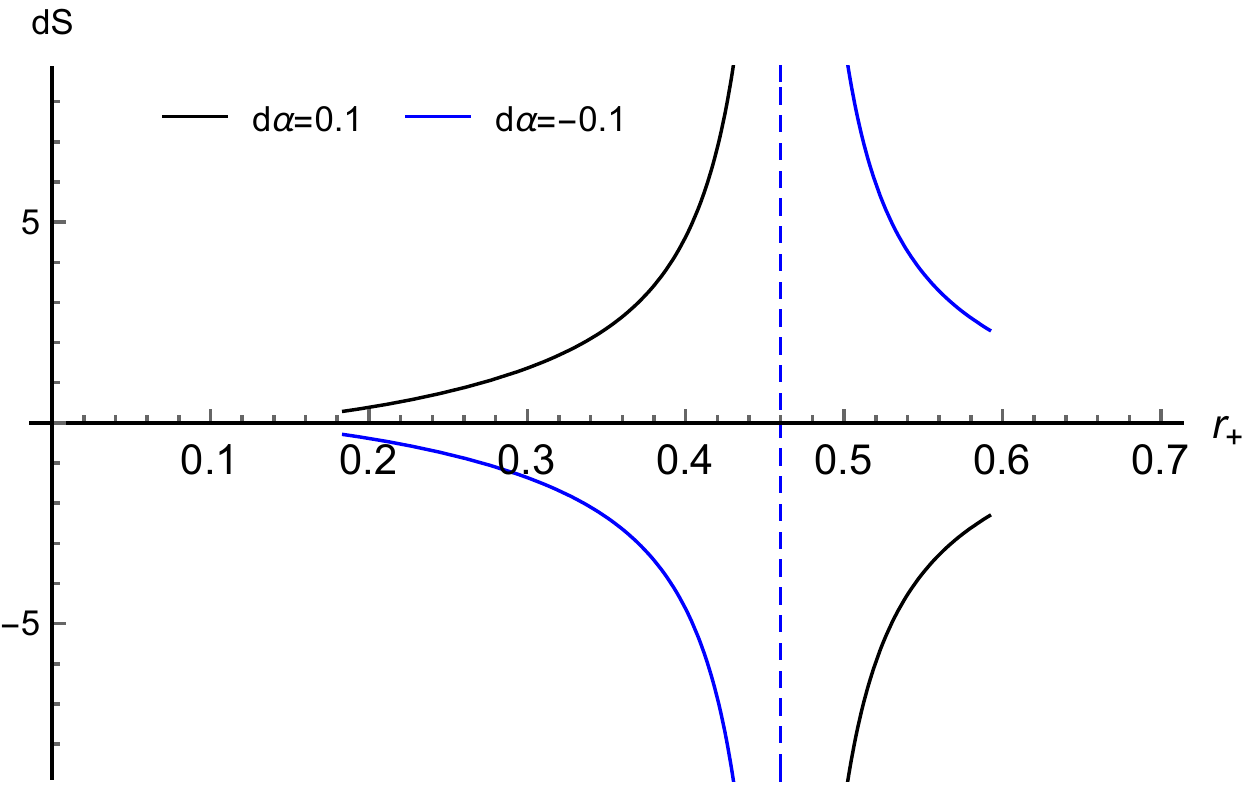}
\caption{The relation between $dS$ and $r_+$ which parameter values are $M = 0.5$, $l = 1$, $q=\omega=0.1$, $\alpha=0.15$ and $dt=0.0001$.}
\label{fig:dS1}
\end{figure}

In the extended phase space, the radius $r$ changes from the initial black hole horizon radius $r_+$ to the changed horizon radius $r_+ + dr_+$ after the scattering of a scalar field. The initial and final states of the black hole are represented by $\left(M,Q,l,\alpha,r_{+}\right)$ and $\left(M+dM,Q+dQ,l+dl,\alpha+d\alpha,r_{+}+dr_{+}\right)$, respectively. They satisfy
\begin{equation}
f\left(M,Q,l,\alpha,r_{+}\right)=f\left(M+dM,Q+dQ,l+dl,\alpha+d\alpha,r_{+}+dr_{+}\right)=0.
\label{eqn:ext fE0}
\end{equation}
The relationship between $f\left(M,Q,l,\alpha,r_{+}\right)$ and $f\left(M+dM,Q+dQ,l+dl,\alpha+d\alpha,r_{+}+dr_{+}\right)$ is
\begin{equation}
\begin{aligned}
&f\left(M+dM,Q+dQ,l+dl,\alpha+d\alpha,r_{+}+dr_{+}\right)=f\left(M,Q,l,\alpha,r_{+}\right)\\
&+\frac{\partial f}{\partial M}|_{r=r_{+}}+\frac{\partial f}{\partial Q}|_{r=r_{+}}+\frac{\partial f}{\partial l}|_{r=r_{+}}+\frac{\partial f}{\partial\alpha}|_{r=r_{+}}+\frac{\partial f}{\partial r}|_{r=r_{+}}.\\
\end{aligned}
\label{eqn:ext T df}
\end{equation}
Substituting Eqs. $\left(\ref{eqn:T Df}\right)$ and $\left(\ref{eqn:ext fE0}\right)$ into Eq. $\left(\ref{eqn:ext T df}\right)$, we can obtain the first law of thermodynamics of the black hole in the extended phase space
\begin{equation}
dM=TdS+\varphi dQ+VdP+\mathcal{A}d\alpha.
\label{eqn:ext T 1st}
\end{equation}
where $\mathcal{A}$ is the conjugate quantity of Gauss-Bonnet parameter and is defined as
\begin{equation}
\mathcal{A}=\left(\frac{\partial M}{\partial\alpha}\right)_{S,Q,P}=\frac{1}{2r_{+}}+2\pi T\left(1-2ln\frac{r_{+}}{\sqrt{\alpha}}\right).
\label{eqn:A}
\end{equation}
Using Eqs. $\left(\ref{eqn:entropy}\right)$, $\left(\ref{eqn:M}\right)$, $\left(\ref{eqn:dUdQ}\right)$, $\left(\ref{eqn:ext T 1st}\right)$ and $\left(\ref{eqn:A}\right)$, we get
\begin{equation}
dS=\frac{\left(1+\frac{2\alpha}{r_{+}^{2}}\right)\left(\omega-q\varphi\right)^{2}r_{+}^{2}dt-\left(1+\frac{2\alpha}{r_{+}^{2}}\right)\left[\frac{1}{2r_{+}}+2\pi T\left(1-2ln\frac{r_{+}}{\sqrt{\alpha}}\right)\right]d\alpha}{\left(1+\frac{2\alpha}{r_{+}^{2}}\right)T-\frac{3r_{+}}{4\pi l^{2}}}.
\label{eqn: ext T dS}
\end{equation}
Based on Eq. $\left(\ref{eqn: ext T dS}\right)$, for large enough $T$, the denominator is positive. On the contrary, for small enough $T$, the denominator is negative. Since $d\alpha$ is arbitrary, the sign of the numerator in Eq. $\left(\ref{eqn: ext T dS}\right)$ is indefinite. It means the second law of thermodynamics is not always satisfied for the extremal or near-extremal black hole. As shown in Fig. \ref{fig:dS1}, we plot the graph of $dS$ with different values of $d\alpha$. We fix $M = 0.5$, $l = 1$, $q=\omega=0.1$, $\alpha=0.15$ and $dt=0.0001$.

\section{Overcharging problem under the scalar field}
\label{sec:wccc}
\begin{figure}[htb]
\begin{center}
\subfigure[{$f\left( r\right)$ in non-extremal black holes.}\label{fig:WCCC1}]{
\includegraphics[width=0.3\textwidth]{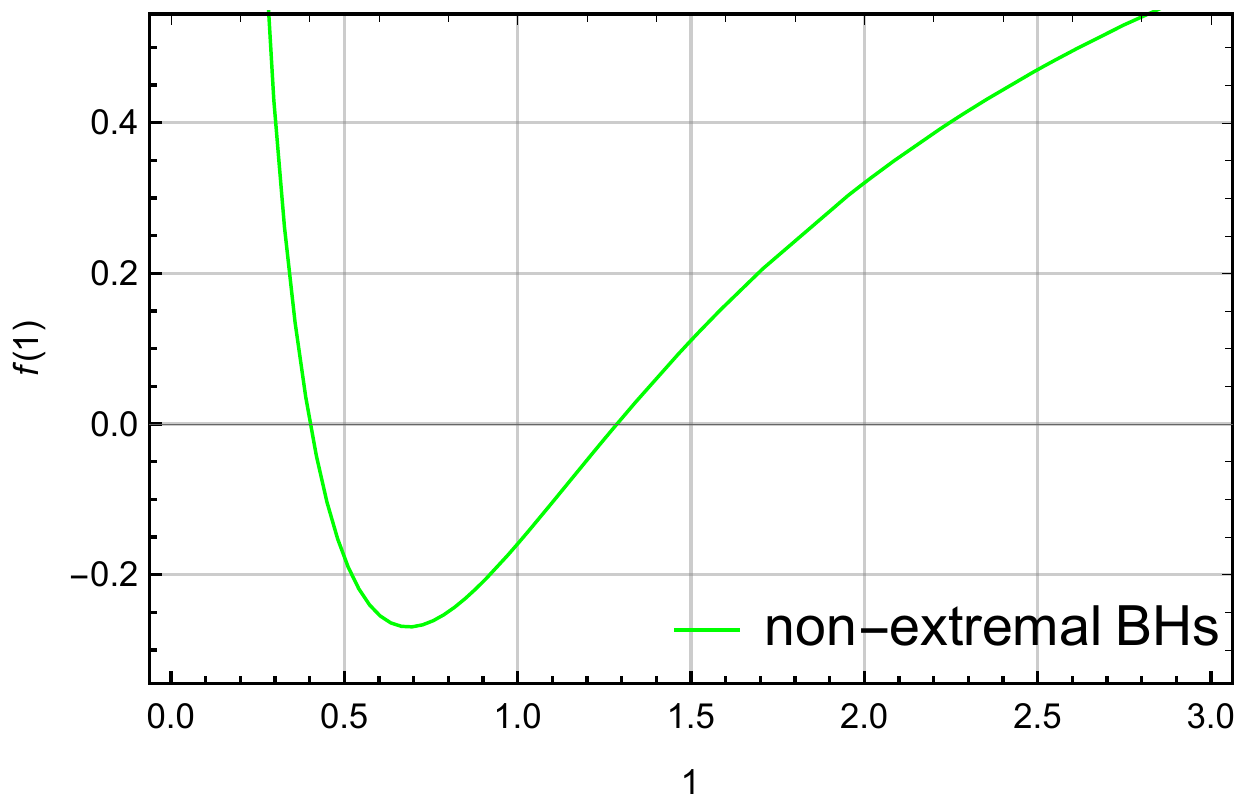}}
\subfigure[{$f\left( r\right)$ in extremal black holes.}\label{fig:WCCC2}]{
\includegraphics[width=0.3\textwidth]{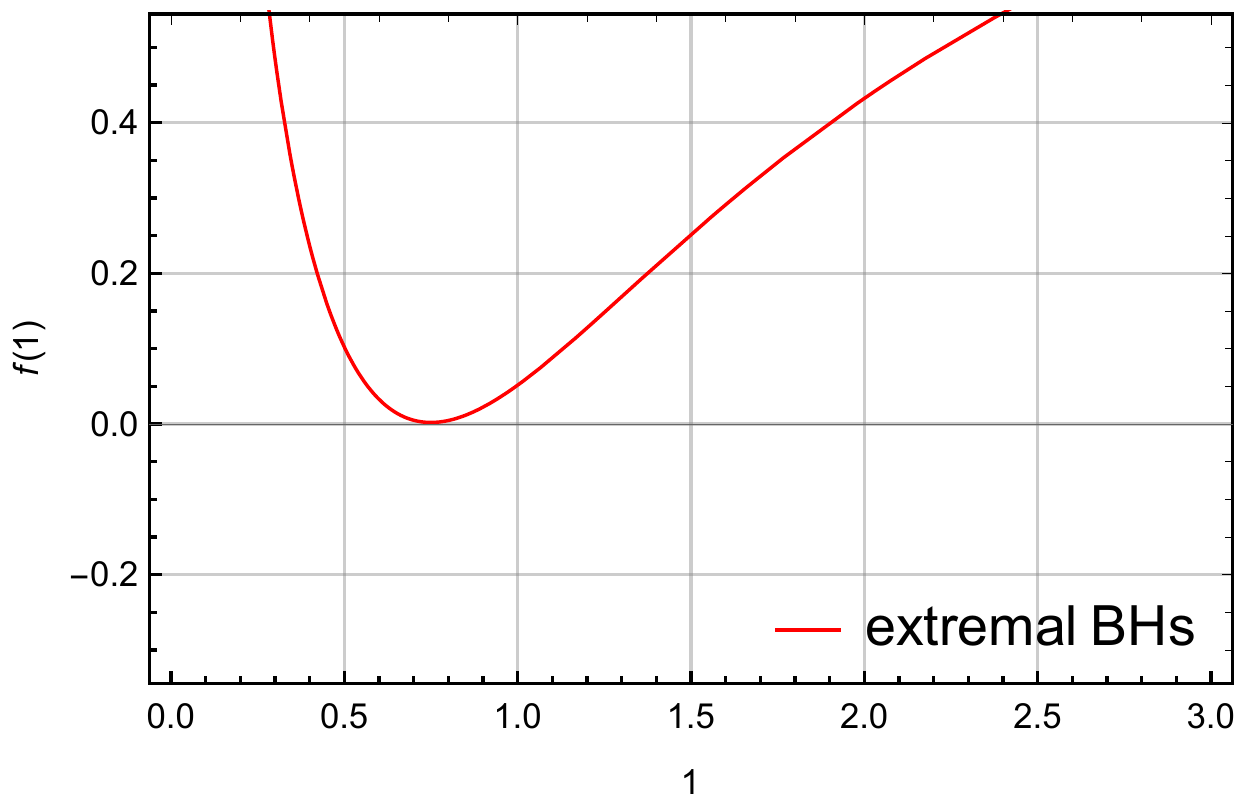}}
\subfigure[{$f\left( r\right)$ in naked singularities.}\label{fig:WCCC3}]{
\includegraphics[width=0.3\textwidth]{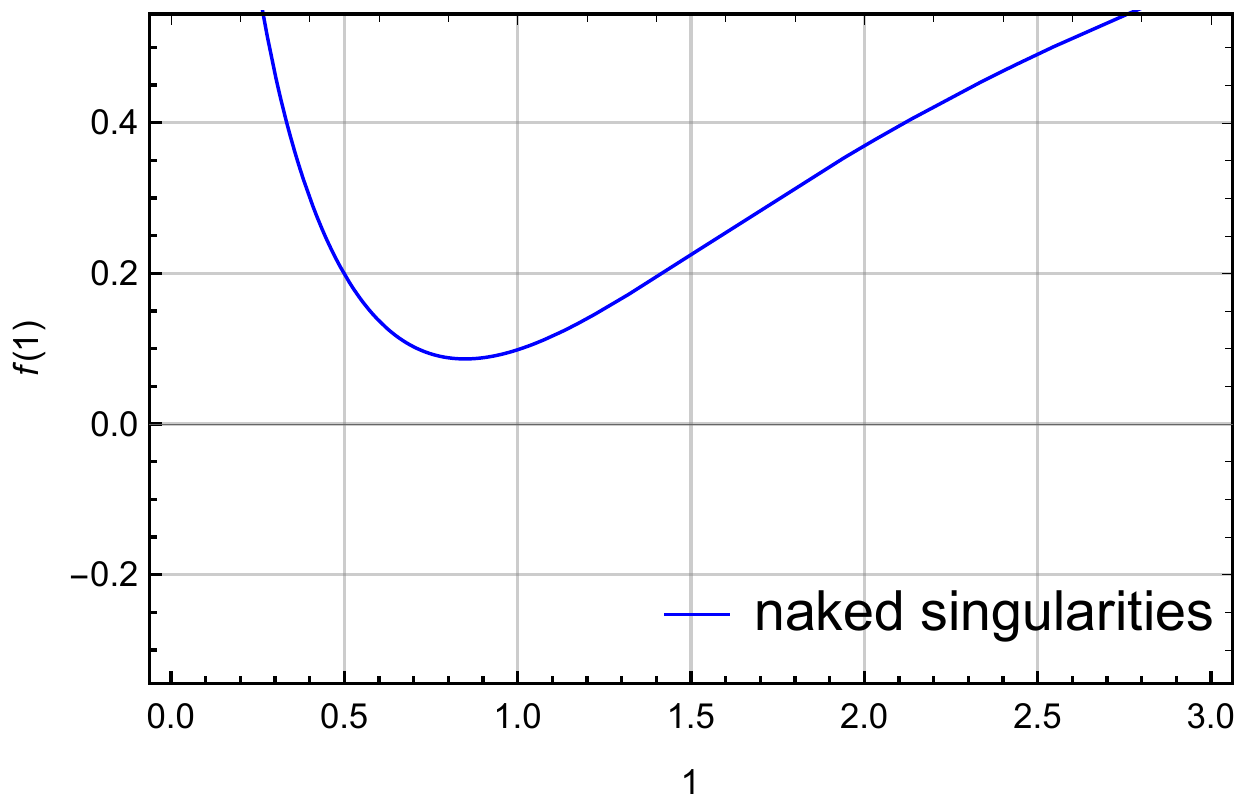}}
\end{center}
\caption{Graphs of $f\left( r\right)$ for given states of the GB-AdS black holes.}%
\label{fig:WCCC}
\end{figure}

In this section, we investigate the validity of the WCCC by the scattering of a scalar field in the normal and extended phase spaces. An effective way to test the validity of WCCC is to check whether the event horizon exists after the scalar field scattering. The event horizon is determined by the function $f(r)$. In the initial state, the minimum value of $f(r)$ is negative or zero and $f(r)=0$ has real roots. It means event horizon exists. After the scalar field scattering, the mass and charge of the black hole change during the infinitesimal time interval $dt$. Besides, the minimum value of $f(r)$ also changes. If the minimum value of $f(r)$ changes to a negative value or zero, as shown in Fig. \ref{fig:WCCC1} and \ref{fig:WCCC2}, the event horizon exists. Then, the black hole cannot be overcharged and the WCCC is effective. Otherwise, as shown in Fig. \ref{fig:WCCC3}, the minimum value of $f(r)$ changes to a positive value. The event horizon doesn't exist. Consequently, the black hole is overcharged and the WCCC is ineffective. Assuming that there is a minimum value of $f(r)$ at $r = r_0$.
For the near-extremal and extremal black holes, the minimum value is not greater than zero
\begin{equation}
\delta\equiv f(r_{0})\leq0.
\end{equation}
where $\delta=0$ corresponds to the extremal black hole.
As before, we first give the following partial derivative formulas
\begin{equation}
\begin{aligned}
&\frac{\partial f}{\partial M}|_{r=r_{0}}=-\frac{2}{r_{0}+\frac{2\alpha}{r_{0}}\left(1-\delta\right)},\\
&\frac{\partial f}{\partial Q}|_{r=r_{0}}=\frac{2\frac{Q}{r_{0}}}{r_{0}+\frac{2\alpha}{r_{0}}\left(1-\delta\right)},\\
&\frac{\partial f}{\partial l}|_{r=r_{0}}=-\frac{2r_{0}^{2}}{l^{3}}\frac{1}{1+\frac{2\alpha}{r_{0}^{2}}\left(1-\delta\right)},\\
&\frac{\partial f}{\partial\alpha}|_{r=r_{0}}=\frac{\left(1-\delta\right)^{2}}{r_{0}^{2}+2\alpha\left(1-\delta\right)},\\
&\frac{\partial f}{\partial r}|_{r=r_{0}}=0.
\end{aligned}
\label{eqn:W Df}
\end{equation}

\subsection{Overcharging problem in the normal phase space}
After the scattering of a scalar field, the physical parameters of the black hole change from the initial
state $(M, Q, r_0)$ to the final state $(M +dM, Q +dQ, r_0 + dr_0)$. At the final state, the value of $f(r)$ at $r=r_0 + dr_0$ satisfies
\begin{equation}
\begin{aligned}
&f\left(M+dM,Q+dQ,r_{0}+dr_{0}\right)=f\left(M,Q,r_{0}\right)\\
&+\frac{\partial f}{\partial M}|_{r=r_{0}}+\frac{\partial f}{\partial Q}|_{r=r_{0}}+\frac{\partial f}{\partial r}|_{r=r_{0}}.\\
\label{eqn:nor W f}
\end{aligned}
\end{equation}
Substituting Eqs. $\left(\ref{eqn:dMdQ}\right)$ and $\left(\ref{eqn:W Df}\right)$ into Eq. $\left(\ref{eqn:nor W f}\right)$, we obtain
\begin{equation}
f\left(M+dM,Q+dQ,r_{0}+dr_{0}\right)=\delta-\frac{2r_{+}^{2}(\omega-q\text{\ensuremath{\varphi}})(\omega-q\frac{Q}{r_{0}})}{r_{0}+\frac{2\alpha}{r_{0}}\left(1-\delta\right)}dt.
\label{eqn:nor W f1}
\end{equation}
When the initial black hole is an extremal black hole, $r_0 = r_+$ and $\delta = 0$. Hence, the minimum value of final
state metric function $f (r)$ takes on the form as
\begin{equation}
f\left(M+dM,Q+dQ,r_{0}+dr_{0}\right)=-\frac{2r_{0}^{2}(\omega-q\text{\ensuremath{\varphi}})^{2}}{r_{0}+\frac{2\alpha}{r_{0}}}dt\leq0.
\end{equation}
Therefore, the black hole can not be overcharged and the WCCC is valid. If $\omega=q\varphi$, the extremal black hole still extremal black hole. If $\omega\neq q\varphi$, the extremal black hole will change to non-extremal black hole.
When it is a near-extremal black hole, $\delta$ is a small negative quantity and $r_{+}$ is numerically greater than $r_{0}$. Eq. $\left(\ref{eqn:nor W f1}\right)$ can be regarded as a quadratic function of $\omega$. When $\omega=\frac{1}{2}qQ\left(\frac{1}{r_{+}}+\frac{1}{r_{0}}\right)$, the function $f\left(M+dM,Q+dQ,r_{0}+dr_{0}\right)$ has the maximum value
\begin{equation}
f\left(M+dM,Q+dQ,r_{0}+dr_{0}\right)_{max}=\delta+\frac{q^{2}Q^{2}(r_{+}-r_{0})^{2}}{2r_{0}^{3}+4\alpha r_{0}^{2}\left(1-\delta\right)}dt.
\label{eqn:nor W f3}
\end{equation}
We define that
\begin{equation}
r_{0}=r_{+}\left(1-\epsilon\right),
\end{equation}
where $0<\epsilon\leqslant1$. Then Eq. $\left(\ref{eqn:nor W f3}\right)$ is rewritten as
\begin{equation}
f\left(M+dM,Q+dQ,r_{0}+dr_{0}\right)_{max}=\delta-\frac{q^{2}Q^{2}\epsilon^{2}}{2r_{+}\left(1-\epsilon\right)^{3}+4\alpha\left(1-\epsilon\right)^{2}\left(1-\delta\right)}dt.
\end{equation}
The second term of the above equation can be ignored since the quantity $\epsilon$ is infinitesimal.
Hence the metric function has a minimum negative value, which indicates that the event horizon also exists in the finial state. The black hole can not be overcharged by the scattering of
the scalar field. Therefore, the WCCC is valid in the near-extremal black hole.

\subsection{Overcharging problem in the extended phase space}
\begin{figure}[h]
\centering
\includegraphics[scale=0.65]{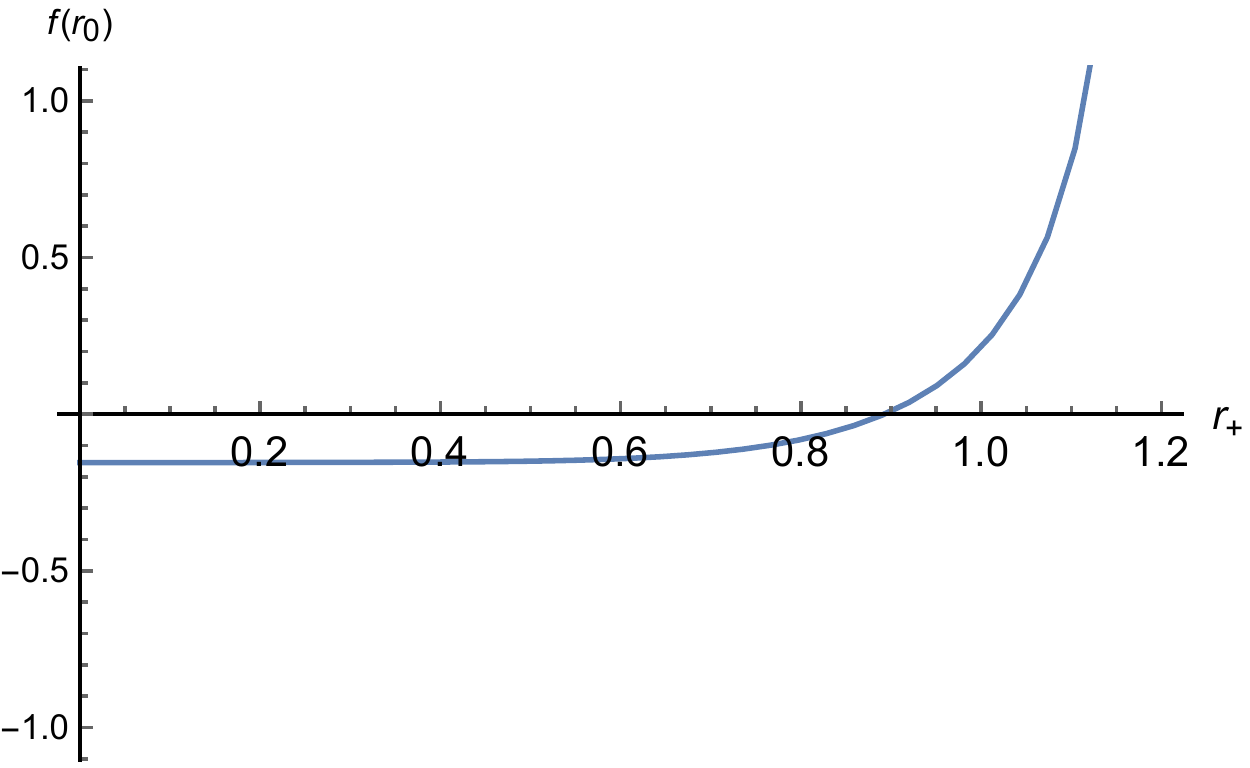}
\caption{The relation between $f(r_0)$ and $r_+$ which parameter values are $\epsilon=0.00000000001, \delta = -0.1,Q=0.7, d\alpha = -0.1, \alpha = 1$ and $l=1$.}
\label{fig:f0}
\end{figure}

In the extended phase space, the physical parameters of the black hole change from the initial
state $\left(M,Q,l,\alpha,r_{0}\right)$ to the final state $\left(M+dM,Q+dQ,l+dl,\alpha+d\alpha,r_{0}+dr_{0}\right)$ after the scattering of a scalar field. At the final state, the value of $f(r)$ at $r=r_0 + dr_0$ satisfies
\begin{equation}
\begin{aligned}
&f\left(M+dM,Q+dQ,l+dl,\alpha+d\alpha,r_{0}+dr_{0}\right)\\
&=f\left(M,Q,l,\alpha,r_{0}\right)+\frac{\partial f}{\partial M}|_{r=r_{0}}+\frac{\partial f}{\partial Q}|_{r=r_{0}}+\frac{\partial f}{\partial l}|_{r=r_{0}}+\frac{\partial f}{\partial\alpha}|_{r=r_{0}}+\frac{\partial f}{\partial r}|_{r=r_{0}}\\
&=\delta-\frac{2TdS}{r_{0}+\frac{2\alpha}{r_{0}}\left(1-\delta\right)}-\frac{2Q}{r_{0}+\frac{2\alpha}{r_{0}}\left(1-\delta\right)}\left(\frac{1}{r_{+}}-\frac{1}{r_{0}}\right)dQ+\frac{2}{r_{0}+\frac{2\alpha}{r_{0}}\left(1-\delta\right)}\left(\frac{r_{+}^{3}}{l^{3}}-\frac{r_{0}^{3}}{l^{3}}\right)dl\\
&-\frac{2}{r_{0}+\frac{2\alpha}{r_{0}}\left(1-\delta\right)}\left[\frac{1}{2r_{+}}+2\pi T\left(1-2ln\frac{r_{+}}{\sqrt{\alpha}}\right)-\frac{\left(1-\delta\right)^{2}}{2r_{0}}\right]d\alpha.\\
\label{eqn:ext W f}
\end{aligned}
\end{equation}
Considering the extremal black hole, which implies $r_{+}=r_{0}$, $T=0$ and $\delta=0$. Therefore, we find that the minimum value of $f (r)$ of the final black hole becomes
\begin{equation}
f\left(M+dM,Q+dQ,l+dl,\alpha+d\alpha,r_{0}+dr_{0}\right)=0
\end{equation}
which means that the extremal black hole is still extremal.
When the initial black hole is a neat-extremal black hole, $r_0$ and $r_+$ do not coincide and $\delta < 0$. As before, we define that
\begin{equation}
r_{0}=r_{+}\left(1-\epsilon\right),
\end{equation}
where $0<\epsilon\leqslant1$. Besides, $f(r_{+})=0 $, $f^{\prime}(r_{+})$ is very close to zero and $f^{\prime}(r_{0})=0$. Then, Eq. $\left(\ref{eqn:ext W f}\right)$ is rewritten as
\begin{equation}
\begin{aligned}
&f\left(M+dM,Q+dQ,l+dl,\alpha+d\alpha,r_{0}+dr_{0}\right)\\
&=\delta-\frac{2T}{r_{+}\left(1-\epsilon\right)+\frac{2\alpha}{r_{+}\left(1-\epsilon\right)}\left(1-\delta\right)}dS\\
&+\frac{\left(1-\delta\right)^{2}-1-4\pi Tr_{+}\left(1-2ln\frac{r_{+}}{\sqrt{\alpha}}\right)}{r_{+}^{2}\left(1-\epsilon\right)^{2}+2\alpha\left(1-\delta\right)}d\alpha\\
&+\frac{\epsilon}{r_{+}^{2}\left(1-\epsilon\right)^{2}+2\alpha\left(1-\delta\right)}\left[2Q+6r_{+}^{4}\left(1-\epsilon\right)\frac{1}{l^{3}}+1+4\pi Tr_{+}\left(1-2ln\frac{r_{+}}{\sqrt{\alpha}}\right)\right]d\alpha .
\end{aligned}
\label{eqn:ext W f2}
\end{equation}
Since the quantity $\epsilon$ is infinitesimal, the forth line of Eq. $\left(\ref{eqn:ext W f2}\right)$ can be ignored. Both terms in the second row are negative. However, the sign of the third line in Eq. $\left(\ref{eqn:ext W f2}\right)$ is  indefinite. As shown in Fig. \ref{fig:f0} , we set $\epsilon=0.00000000001, \delta = -0.1, Q=0.7, d\alpha = -0.1, \alpha = 1$ and $l=1$. From the figure we can know that the sign of $f\left(M+dM,Q+dQ,l+dl,\alpha+d\alpha,r_{0}+dr_{0}\right)$ will change from negative to positive when $r_0$ increase. Therefore, in the extended phase space, the near-extremal black hole can be overcharged and the effectiveness of WCCC is indefinite.
\section{Discussion and Conclusion}
\label{sec:con}
In this paper, we investigated the thermodynamics and overcharging problem in a 4D Gauss-Bonnet AdS black hole under the scalar field. First, we reviewed the solutions of the 4D Gauss-Bonnet AdS black hole. Moreover, the variations of this black hole's energy and charge within the infinitesimal time interval are investigated. Then we tested the validity of the first and second laws of thermodynamics and the weak cosmic censorship conjectures. It is worth noting that the RN-AdS black hole solution will be recovered when we take the limit $\alpha\rightarrow0$. In Table \ref{tab:wccc1} and Table \ref{tab:wccc2}, we summarize and compare the results of thermodynamics and WCCC both for the 4D Gauss-Bonnet AdS black hole and RN-AdS black hole. Moreover, the detailed formulas of the first law of thermodynamics for the 4D Gauss-Bonnet AdS black hole and the RN-AdS black hole are shown in Table \ref{tab:1st}.
\begin{table}[htb]
\begin{centering}
\begin{tabular}{|p{0.7in}|c|c|p{3in}|}
  \hline &\multicolumn{3}{|c|}{Normal phase space} \\
  \hline &1st law &2nd law &WCCC \\
  \hline GB-AdS black hole  &Satisfied  &Satisfied  &Satisfied in the extremal black hole and near-extremal black hole. After the scalar field scattering, the extremal black hole will change to a non-extremal black hole under certain conditions.\\
  \hline RN-AdS black hole  &Satisfied  &Satisfied  &Satisfied for the extremal and near-extremal black holes. After the
scalar field scattering, the extremal black hole will change to a non-extremal black hole under certain conditions. \\
  \hline
\end{tabular}
\par\end{centering}
\caption{{\footnotesize{}{}{}{}Results for the first and second laws of
thermodynamics and the overcharging problem in the normal phase space.}}
\label{tab:wccc1}
\end{table}

\begin{table}[htb]
\begin{centering}
\begin{tabular}{|p{0.7in}|c|c|p{3in}|}
  \hline  &\multicolumn{3}{|c|}{Extended phase space}\\
  \hline &1st law &2nd law &WCCC\\
  \hline GB-AdS black hole  &Satisfied &Indefinite &Satisfied for the extremal black hole and indefinite for the near-extremal black hole. After the scalar field scattering, the extremal black hole stays extremal. \\
  \hline RN-AdS black hole  &Satisfied  &Indefinite &Satisfied for the extremal and near-extremal black holes. After the scalar field scattering, the extremal black hole stays extremal. \\
  \hline
\end{tabular}
\par\end{centering}
\caption{{\footnotesize{}{}{}{}Results for the first and second laws of
thermodynamics and the overcharging problem in the extended phase space.}}
\label{tab:wccc2}
\end{table}

As shown in Table \ref{tab:wccc1} and Table \ref{tab:wccc2}, for the 4D Gauss-Bonnet AdS black hole, the first law of thermodynamics is satisfied both in the normal and extended phase spaces after the scattering of a scalar field.
\begin{table}[htb]
\begin{centering}
\begin{tabular}{|p{2.0in}|p{4.0in}|}
\hline
Types of black holes & 1st law in the extended phase space  \tabularnewline
\hline
GB-AdS BH  & $dM=TdS+\varphi dQ+VdP+\left[\frac{1}{2r_{+}}+2\pi T\left(1-2ln\frac{r_{+}}{\sqrt{\alpha}}\right)\right]d\alpha.$\tabularnewline
\hline
RN-AdS BH   & $dM=TdS+\text{\ensuremath{\varphi}}dQ+VdP$.\tabularnewline
\hline
\end{tabular}
\par\end{centering}
\caption{{\footnotesize{}{}{}{}Results for the first
thermodynamic law under different conditions in the extended phase space.}}
\label{tab:1st}
\end{table}
However, the second law of thermodynamics are not the same in this two cases. In the normal phase space, the second law of thermodynamics is satisfied. In the extended phase space, the validity of the second law of thermodynamics is indefinite. Furthermore, the overcharging problem is considered both in the normal and extremal phase spaces. In the normal phase space, the black hole cannot be overcharged after the scattering of a scalar field and the WCCC is satisfied. In the extended phase space, the extremal black hole cannot be overcharged. However, the near-extremal black hole can be overcharged under certain conditions. Therefore, the validity of the WCCC is indefinite.

In Ref. \cite{intro-Ying:2020bch}, thermodynamics and overcharging problem with pressure and volume of 4D Gauss-Bonnet AdS black holes are discussed by the charged particle absorption. In Ref. \cite{intro-Zeng:2019hux}, weak cosmic censorship conjecture with pressure and volume in the Gauss-Bonnet AdS black hole ($d>4$) is discussed. It is hoped that more methods can be used to study the thermodynamic properties of 4D Gauss-Bonnet AdS black holes in the future. This will prompt us to further explore the deep relationship between the thermodynamic properties of the new four-dimensional black hole and its boundary conditions, and to learn more about the black hole.
\begin{acknowledgments}
We are grateful to Deyou Chen, Peng Wang, Haitang Yang and Jun Tao for useful discussions. This work is supported in part by NSFC (Grant No. 11747171), Xinglin Scholar Research Premotion Project of Chengdu University of TCM(Grants nos. QNXZ2018050, ZRYY1729 and ZRYY1921),the key fund project for Education Department of Sichuan (Grant no. 18ZA0173), the Special Talent Projects of Chizhou University (Grant no. RZ2000000591).
\end{acknowledgments}

\end{document}